# ELECTRON CLOUD MEASUREMENTS IN FERMILAB BOOSTER*

S.A.K. Wijethunga†, N. Eddy, J. Eldred, C.Y. Tan, B. Fellenz, E. Pozdeyev, R.V. Sharankova
Fermi National Accelerator Laboratory, Batavia, Illinois 60510.

*Abstract*

Fermilab Booster synchrotron requires an intensity upgrade from $4.5\times10^{12}$ to $6.5\times10^{12}$ protons per pulse as a part of Fermilab's Proton Improvement Plan-II (PIP-II). One of the factors which may limit the high-intensity performance is the fast transverse instabilities caused by electron cloud effects. According to the experience in the Recycler, the electron cloud gradually builds up over multiple turns inside the combined function magnets and can reach final intensities orders of magnitude greater than in a pure dipole. Since the Booster synchrotron also incorporates combined function magnets, it is important to measure the presence of electron cloud. The presence or apparent absence of the electron cloud was investigated using two different methods: measuring bunch-by-bunch tune shift by changing the bunch train structure at different intensities and propagating a microwave carrier signal through the beampipe and analyzing the phase modulation of the signal. This paper presents the results of the two methods and corresponding simulation results conducted using PyECLOUD software.

## INTRODUCTION

The formation of an electron cloud (EC) can severely limit the performance of high-intensity proton accelerators due to transverse instabilities, transverse emittance growth, particle losses, vacuum degradation, heating of the chamber's surface, etc. [1-4]. Studies conducted previously at the Fermilab Recycler facility have shown that combined function magnets can trap the EC due to its magnetic mirror effect. According to their simulations, EC accumulates over many revolutions inside a combined function magnet and can reach final intensities orders of magnitudes higher than inside a pure dipole [3].

In order to meet the demands of Fermilab's Proton Improvement Plan-II (PIP-II), the Fermilab Booster [5], a rapid-cycling (15 Hz) synchrotron which is equipped with 96 combined function magnets, will need to deliver a high-intensity beam of $6.5\times10^{12}$ protons per pulse, representing a 44% increase in current intensity [6]. Therefore, it is important to find evidence of the presence of an EC in the PIP-II era Booster and evaluate if it has any impact on the desired performance.

Over the years, several techniques have been developed to measure the EC in accelerators [4, 7-11]. This paper presents two such techniques employed to explore the existence or absence of the EC in the Fermilab Booster. These techniques include measuring the bunch-by-bunch tune shift by changing the bunch train structure at different intensities and propagating a microwave carrier signal through the beampipe and analyzing the phase modulation of the signal.

## BUNCH-BY-BUNCH TUNE SHIFT

In order to trap electrons in a magnetic field, it is necessary to have a train of closely spaced bunches [3]. In the absence of a following bunch, these secondary electrons can go through a few elastic reflections before being absorbed by the vacuum chamber. Hence, a single bunch, *i.e.*, a clearing bunch, following the main batch can be used to clear the EC as it kicks the electrons into the vacuum chamber. Since an EC act like a lens providing additional focusing or defocusing to the beam, the clearing of the EC can be observed as a shift of the betatron tune.

The existence of the EC in the Booster was first investigated by introducing different gaps in the bunch structure with varying beam intensities. Then the corresponding tune shifts were analyzed. This paper presents measurements taken with vertical pings for four different beam intensities protons per pulse (ppp): $1.7\times10^{12}$, $4.5\times10^{12}$, $5.0\times10^{12}$, and $5.5\times10^{12}$ near transition. According to both simulation and theory, the strongest EC in the Booster is expected to occur near transition when the bunch length is at its shortest. Two different bunch structures were created by misaligning the laser notcher and the notcher kicker [12,13].

Figure 1 presents the vertical tune variation of each bunch at a turn near transition for both bunch structures and four intensities. Bunches with partial intensity due to the laser notcher show a positive shift compared to the rest of the beam. However, this may be due to the impedance tune depression, as the low-intensity beam shows a smaller shift compared to the high-intensity beam. For the other bunches, no significant difference between the opposite notch beam and the nominal beam.

Figure 2 depicts the vertical tune shifts of the opposite notch bunch structure with respect to the nominal notch bunch structure of the last ~2500 turns before the transition for the four intensities. Note that the tune difference of each turn was calculated by considering the bunches with typical tunes (unaffected by the nearby notches) and taking the average of them. According to Ref. [3], a positive tune shift in the horizontal direction indicates the presence of an EC at the beam center, and a negative tune shift in the vertical direction indicates the maximum density of the EC near the walls of the vacuum chamber. Adding a clearing bunch can reduce these tune shifts. However, the results do not show either significant or consistent tune shift near transition for all four intensities.

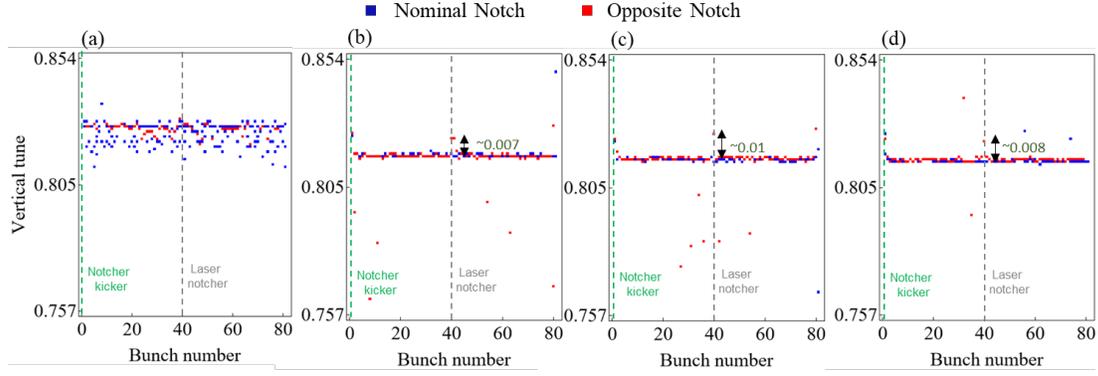

Figure 1: Bunch tune variation near transition for nominal and opposite notches for the four different beam intensities (ppp): (a) $1.7\times10^{12}$, (b) $4.5\times10^{12}$, (c) $5.0\times10^{12}$, and (d) $5.5\times10^{12}$. The dashed line indicates the location of the notcher kicker (green), and laser notcher (light gray).

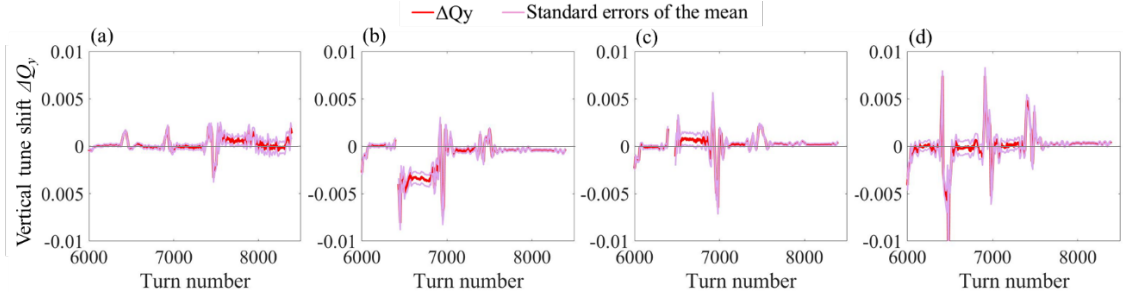

Figure 2: Vertical tune shift due to the gap near transition for the four different beam intensities (ppp): (a) $1.7\times10^{12}$, (b) $4.5\times10^{12}$, (c) $5.0\times10^{12}$, and (d) $5.5\times10^{12}$. The error was calculated by taking the standard errors of the mean. The discontinuity in some plots is due to the distorted tune bands.

## SIMULATIONS

PyECLOUD code was used to simulate the EC build-up inside a combined function magnetic located in the Booster synchrotron [14]. Table 1 lists the main input parameters used in the simulations. The cross-section of the combined function magnet was considered a rectangle with diploe and quadrupole magnetic fields.

Table 1: Input parameters in PyECLOUD simulations.

| Parameter | Transition | Injection |
|---|---|---|
| Beam energy [GeV] | 4.2 | 0.4 |
| Bunch spacing [ns] | 19.2 | 26.4 |
| Bunch length, $\sigma$ [m] | 0.25 | 0.57 |
| SEY, $\delta$ | 1.8 | |

The simulation was conducted for 3 turns near injection and transition for the same intensities as the measurements (ppp): $1.7\times10^{12}$, $4.5\times10^{12}$, $5.0\times10^{12}$, $5.5\times10^{12}$ and additionally, PIP-II intended intensity $6.5\times10^{12}$. PyEcloud did not show meaningful EC accumulation near injection with 1.8 SEY and 0.57 m (rms) bunch length. Figure 3 shows the simulations near transition.

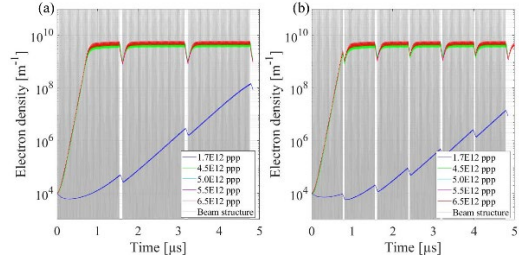

Figure 3: EC build-up for different beam intensities (a) nominal notch and (b) opposite notch near transition.

Based on the plots shown above, it can be observed that EC exists within Booster. The accumulation of EC in opposite notch bunch structures is slower for low-intensity beam (ppp): $1.7\times10^{12}$ compared to the nominal bunch structure. Nevertheless, the EC saturation is almost identical for all the high-intensity beams, irrespective of their bunch structure, *i.e.* these simulations show that the notch configurations that were used minimally clears EC.

The measurements and simulations were compared to a simple theoretical model that gives the relation between the tune shift $\Delta Q$ and the corresponding EC density $\rho$ [15,16],

$$\Delta Q = \frac{r_p}{\gamma \beta^2} \langle \beta \rangle \rho C \frac{x^2}{(x+y)^2} \qquad (1)$$

where, $r_p$ is the classical proton radius, $C$ is the circumference, $\langle\beta\rangle$ is the average beta function, $\beta$ is the relativistic beta, $\gamma$ is the Lorentz factor, and $x$ and $y$ are the semi-aperture dimensions. According to the calculations, PyECLOUD simulations show the maximum tune shift near transition of about 0.001 (with 1.8 SEY), which is considerably lower than what we have observed in measurements.

## MICROWAVE MEASUREMENTS

Our second method involved transmitting a microwave carrier signal through the beampipe and examining the phase modulation caused by the EC. This method can be applied in two distinct variations: transmission and resonance [17]. The transmission method required a long section of a beampipe with a uniform cross-section to avoid reflections which can lead to undefined propagation lengths. Hence, we chose the resonance TE wave method [7,8].

In order to couple microwaves in and out of the beampipe, beam position monitor (BPM) systems were used. After taking $S_{21}$ measurements at a few BPM locations (using both horizontal plates and both vertical plates) along the Booster ring, Short 15 (S15) was selected as it had a resonance at 1.355 GHz, which is also close to the pipe's cut-off frequency. At this frequency, the resonator would have an effective length of about 10 m.

Figure 4 illustrates the schematic diagram of the setup used for this experiment. The standing waves are expected to be set up in between the ion pumps, creating a resonant cavity. The phase demodulator box basically consists of a mixer and low pass filter that was specially built for this experiment to measure the phase delay between the generated and the received signals.

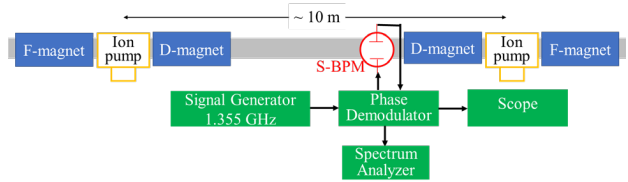

Figure 4: Schematic diagram of the experimental setup.

The measurements were carried out with two different beam intensities (ppp): $4.9 \times 10^{12}$ and $3.8 \times 10^{12}$ near transition and near extraction. Ideally, near transition, the bunch length reduces to its minimum, leading to a significant phase shift caused by the maximum EC density. However, the frequency sweep of the Booster introduces complications to this measurement. Hence, we repeated the measurement near extraction because the RF frequency is nearly constant here. The carrier frequency was set to 1.355 GHz and the amplitude to 10 dB.

Figure 5 shows the phase shift of the carrier with the beam, the carrier without the beam, and the beam structure near transition. Figure 6 shows the same results near the extraction. The data presented for intensity $3.8 \times 10^{12}$ is averaged over 512 beam cycles, and intensity $4.9 \times 10^{12}$ is averaged over 256 beam cycles.

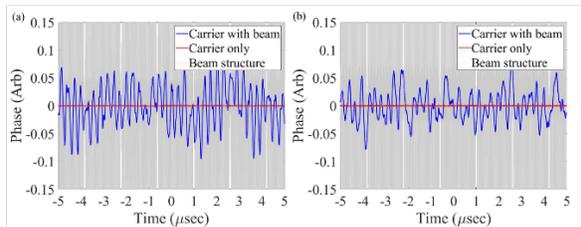

Figure 5: Phase shift near transition for two different beam intensities (ppp): (a) $4.9 \times 10^{12}$ and (b) $3.8 \times 10^{12}$.

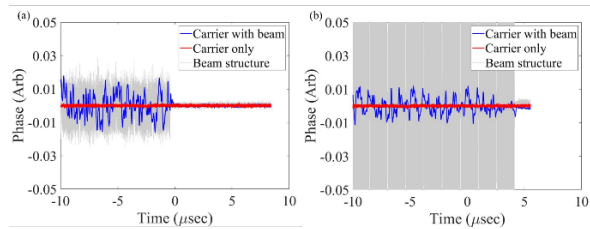

Figure 6: Phase shift near extraction for two different beam intensities (ppp): (a) $4.9 \times 10^{12}$ and (b) $3.8 \times 10^{12}$.

These results show that there is no detectable phase shift due to the beam near transition or extraction for both $4.9 \times 10^{12}$ and $3.8 \times 10^{12}$ (ppp) intensities, and therefore no EC was measured. Further, there is no visible phase shift due to the gaps between the turns.

## CONCLUSION

The presence or absence of the electron cloud (EC) in the PIP-II era Booster was investigated using two techniques, and PyECLOUD software simulations. The bunch-by-bunch tune shift near transition showed a larger tune shift which was not consistent with the simulation's calculated tune shift, indicating the possibility of impedance tune depression. The average tune comparison near the transition did not provide a clear indication of the presence of EC. Furthermore, microwave measurements also showed that there is no detectable EC near both transition and extraction. Both bunch-by-bunch tune shift method and simulations agreed that the gap between the turns is not large enough to clear the EC. Hence the resulting tune shift is not sensitive enough to make any predictions about EC in Booster.

To explore the impact of beampipe conditioning on EC, the microwave measurements will be repeated following the Booster shut-down period in the future. Additionally, SEY measurements of the laminations and epoxy used to build the combined function magnets will be used for more accurate simulation results.

## ACKNOWLEDGMENT

We are grateful to Salah J Chaurize and Kent Triplett for their insights and support throughout these experiments.